\documentclass[10pt,journal,compsoc]{IEEEtran}
\ifCLASSOPTIONcompsoc
  \usepackage[nocompress]{cite}
\else
  \usepackage{cite}
\fi
\ifCLASSINFOpdf
\else
\fi
\usepackage{helvet} 
\usepackage{courier}  
\usepackage[hyphens]{url}  
\usepackage{graphicx} 
\usepackage{color}
\usepackage[utf8]{inputenc}
\usepackage[english]{babel}
\usepackage{wrapfig}
\usepackage{subfigure}
\usepackage{algorithm}
\usepackage{amssymb}
\usepackage{amsmath}
\usepackage[noend]{algpseudocode}
\usepackage{booktabs}
\usepackage{multirow}
\usepackage{wrapfig,booktabs}
\usepackage[T1]{fontenc}

\usepackage[utf8]{inputenc}
\usepackage[utf8]{inputenc}
\usepackage[english]{babel}

\newtheorem{corollary}{Corollary}

\usepackage{ragged2e}
\usepackage{pgfplots}
\usepackage{ltablex}
\usetikzlibrary{positioning}
\usetikzlibrary{fit}
\usetikzlibrary{backgrounds}
\usetikzlibrary{calc}
\usetikzlibrary{shapes}
\usetikzlibrary{mindmap}
\usetikzlibrary{decorations.text}
\pgfplotsset{compat=1.7}
\usepackage{multirow}
\usepackage{siunitx} 
\usepackage{etoolbox} 
\robustify\bfseries

\usepackage{booktabs}
\usepackage{balance}
\usepackage{tikz}
\usetikzlibrary{matrix,positioning}
\usetikzlibrary{arrows.meta,shapes.arrows,chains,decorations.pathreplacing}
\usetikzlibrary{automata}
\usetikzlibrary{calc}
\usepackage{caption}
\usetikzlibrary{decorations.pathreplacing,matrix}

\begin{document}
%
\title{Multi-task Learning for Influence Estimation and Maximization}
%
%
%
%

\author{George~Panagopoulos, Fragkiskos~D.~Malliaros, and Michalis~Vazirgiannis
\IEEEcompsocitemizethanks{\IEEEcompsocthanksitem G. Panagopoulos is with the Computer Science Laboratory (LIX), École Polytechnique, Institute Polytechnique de Paris,  91120 Palaiseau, France \protect\\
E-mail: george.panagopoulos@polytechnique.edu
\IEEEcompsocthanksitem  F. D. Malliaros is with Paris-Saclay University, CentraleSupélec, Inria, Centre for Visual Computing (CVN), 91190 Gif-Sur-Yvette, France \protect\\
E-mail: fragkiskos.malliaros@centralesupelec.fr
\IEEEcompsocthanksitem M. Vazirgiannis is with the Computer Science Laboratory (LIX), École Polytechnique, Institute Polytechnique de Paris,  91120 Palaiseau, France and with the Department of Informatics, Athens University of Economics and Business, 10434 Athens, Greece \protect\\
E-mail: mvazirg@lix.polytechnique.fr
}
\thanks{Manuscript received XXX; revised XXX.}}

\IEEEtitleabstractindextext{%
\begin{abstract}
We address the problem of influence maximization when the social network is accompanied by diffusion cascades.
In the literature, such information is used to compute influence probabilities, which is utilized by stochastic diffusion models in influence maximization. 
Motivated by the recent criticism on diffusion models and the galloping advancements in influence learning, we propose \textsc{IMINFECTOR} (Influence Maximization with INFluencer vECTORs), a method that uses representations learned from diffusion cascades to perform model-independent influence maximization. 
The first part of our methodology is a multi-task neural network that learns embeddings of nodes that initiate cascades (influencer vectors) and embeddings of nodes that participate in them (susceptible vectors). The norm of an influencer vector captures a node's aptitude to initiate lengthy cascades and is used to reduce the number of candidate seeds. The combination of influencer and susceptible vectors form the diffusion probabilities between nodes. These are used to reformulate the computation of the influence spread and propose a greedy solution to influence maximization that retains the theoretical guarantees.
We apply our method in three sizable datasets and evaluate it using cascades from future time steps. \textsc{IMINFECTOR}'s scalability and accuracy outperform various competitive algorithms and metrics from the diverse landscape of influence maximization.
\end{abstract}

\begin{IEEEkeywords}
influence maximization, influence learning, multi-task learning
\end{IEEEkeywords}}

\maketitle

\IEEEdisplaynontitleabstractindextext

%
\IEEEpeerreviewmaketitle

\IEEEraisesectionheading{
\section{Introduction}\label{sec:introduction}}
\IEEEPARstart{S}{ocial} networks have come to play a substantial role in numerous economical and political events, exhibiting their effect on shaping public opinion.
The core component of their potential in micro level is how a user influences another user online, which is depicted in the users' connections and activity. 
Formally, social influence is defined as a directed measure between two users and represents how possible is for the target user to adapt the behavior or copy the action of the source user.
In viral marketing, influence is used to simulate how information flows through the network and find the optimal set of nodes to start a campaign from, the well-known influence maximization (IM) problem.
In IM, the users are represented as nodes and the edges depict a relationship, like following, friendship, coauthorship, etc. A stochastic diffusion model, such as the independent cascade \cite{kempe2003maximizing}, governs how an epidemic traverses the users based on their connections.
It is used to compute the number of users activated during a diffusion simulation, which is called the \textit{influence spread}.
The aim is to find the optimal set of $k$ users that would maximize the influence spread of a diffusion cascade starting from them \cite{kempe2003maximizing}.

One of the main problems in the IM literature is that diffusion models utilize random or uniform influence parameters. Experiments have shown that such approach produces a less realistic influence spread compared to models with empirical parameters \cite{aral2018social}. 
Even when the parameters are learned in an empirical manner from historical logs of diffusion cascades, the independent cascade (IC) model is utilized \cite{continest,bourigault2016representation,saito2009learning}, which is problematic for two reasons.
Apart from severe overfitting due to the massive number of parameters, this approach assumes influence independence throughout related nodes or edges of the same node. This assumption overlooks the network's assortativity, meaning that an influential node is more prone to effect a susceptible node then a less influential node, even when the edge of the latter to the susceptible is stronger than the edge of the former. This effect comes in contrast with the actual mechanics of influence \cite{aral2012identifying}.
In addition, diffusion models themselves suffer from oversimplifying assumptions overlooking several characteristics of real cascades \cite{complex2018} which can provide inaccurate estimations compared to actual cascades \cite{maria2016} while they are highly sensitive to their parameters \cite{du2014influence}. 

To address these issues, we propose a method that learns influencer and susceptible embeddings from cascades, and uses them to perform IM without the use of a diffusion model.
The first part of our method is INFECTOR (INfluencer Vectors), a multi-task neural network that learns influencer vectors for nodes that initiate cascades and susceptible vectors for those that participate in them. 
Our focus on the cascades' initiators is justified from empirical evidence on the predominantly initiating tendency of influencers and it is considerably faster compared to previous influence learning (IL) models \cite{feng2018inf2vec}.
INFECTOR additionally embeds the aptitude of an influencer to create sizable cascades in the norm of her embedding. From this, we derive an estimate of influencers' spread and use it to diminish the number of the candidate seeds for IM. 
Following suit from recent model-independent algorithms \cite{lagree2018algorithms,vaswani2017model}, we overlook the diffusion model and connect each candidate seed (influencer) and every susceptible node with a diffusion probability using the dot product of their respective embeddings, forming a bipartite network. Apart from removing the time-consuming simulations, diffusion probabilities have the advantage of capturing higher-order correlations that diffusion models fail to due to their Markovian nature. Finally, we propose \textsc{IMINFECTOR}, a scalable greedy algorithm that uses a submodular influence spread to compute a seed set, retaining the theoretical guarantee of $1-1/e$.

To evaluate the performance of the algorithm, the quality of the produced seed set is determined by a set of unseen cascades from future time steps, similar to a train and test split in machine learning. We deem this evaluation strategy more reliable than traditional evaluations based on simulations because it relies on actual traces of influence. 
\textsc{IMINFECTOR} outperforms previous IM methods with learned edge weights, either in quality or speed in three sizable networks with cascades.
This is the first time, to the best of our knowledge, that representation learning is used for IM with promising results. 

Our contributions can be summarized as follows: 
\begin{itemize}
    \item \textsc{INFECTOR}: a multi-task learning neural network that captures simultaneously the influence between nodes and the aptitude of a node to create massive cascades. 
    \item \textsc{IMINFECTOR}: a model-independent algorithm that uses the combinations and the norms of the learnt representations to produce a seed set for IM.
    \item A novel, thorough experimental framework with new large scale datasets, seed set size adapted to the network scale and evaluation based on ground truth cascades instead of simulations.
\end{itemize}
\noindent \textbf{Reproducibility}: The implementation and links to the datasets can be found online on github\footnote{https://github.com/geopanag/IMINFECTOR}.

\vspace{1.3cm}
\IEEEraisesectionheading{
\section{Related Work}\label{sec:related}}
The basic solution in the problem of IM is a greedy algorithm that builds a seed set by choosing the node that provides the best marginal gain at each iteration, i.e., the maximum increase of the set's influence spread \cite{kempe2003maximizing}.
The algorithm estimates the influence spread using the live-edge model, a Monte Carlo sampling of edges that estimates the result of simulating a diffusion model.  The algorithm is guaranteed to reach a near optimal solution because of the submodularity if the marginal gain. Having said that, the 1,000 samplings needed for the sufficient estimation of the influence spread for each candidate seed in each iteration, renders the approach infeasible for real world networks. 
Several methods have been developed that achieve notable acceleration and retain the theoretical guarantees using
sketches \cite{sketch} or reverse reachable sets \cite{tim,imm}. Moreover numerous heuristics have been proposed that, though lacking guarantees, exhibit remarkable success in practice \cite{simpath,chen2010scalable}.As the problem of IM became more popular, several methods were developed to derive more realistic versions by adding temporal \cite{liu2013influence}, topic \cite{li2018influence}  and location constraints \cite{zhang2020geodemographic}.
From the perspective of network immunization, the required seed set can be retrieved by finding the must pivotal nodes using optimal percolation \cite{pei2020influencer}. In addition, if the outbreak has already started, a dynamic curing policy based on the graph's cutwidth can be used to control the epidemic \cite{drakopoulos2014efficient}. Such methods are developed for epidemic containment and have not been utilized in the context of IM yet, hence are omitted in this paper.

The main problem of the aforementioned IM approaches is that the established influence spread may lead to seed sets of poor quality. This can be caused either by the assignment of simplistic influence weights or by the diffusion model's innate assumptions.
To address the issue with the random influence weights, some novel influence-maximization approaches assume multiple rounds of influence maximization can take place over the graph, hence multiple simulations can be used to compute the influence probabilities while balancing between the number of nodes influenced in each round and learning influence for non examined parts of the network. Since this is an inherently exploration-exploitation problem these models are based on multi-armed bandits to use the feedback and update of the parameters \cite{wen2017online,sun2018multi,wu2019factorization}.
Though useful, these algorithms are built based on the diffusion models as well, hence they share their aforementioned deficiencies.
Recently more model-independent online learning \cite{lagree2018algorithms} approaches that utilize regret functions without the use of diffusion models have been proposed \cite{vaswani2017model}, but they suffer from important scalability issues and would not be able to scale in real-world social networks, such as the datasets we examine in this work.

Though influence maximization has been framed as a supervised problem with node representations in the past \cite{ivanov2018learning}, representations learnt from actual influence traces have not been broadly utilized. To be specific, there have been few attempts to address influence maximization with learned influence parameters from real past cascades, which serve as benchmark comparisons for our proposed approach \cite{credit,goyal2010learning}. 
These models suffer from overfitting due the number of parameters which is proportional to the number of edges. 
To reduce the number of parameters, a more practical approach is to express the influence probabilities as a combination of the nodes' influence and susceptibility embeddings and learn them from cascades \cite{bourigault2016representation}.
This method however relies on IC, and consequently suffers from the aforementioned influence independence assumption. 
More recent IL methods are devoid of diffusion models and learn the embeddings based on co-occurrence in cascades \cite{feng2018inf2vec}, similarly to node representation learning. This type of IL has not been used yet for IM.
Although more accurate in diffusion prediction, the input of this model consists of node-context pairs derived from the propagation network of the cascade, a realization of the underlying network (e.g., follow edges) based on time-precedence in the observed cascade (e.g., retweets).
This requires looping through each node in the cascade and iterating over the subsequent nodes to search for a directed edge in the network, which has a complexity of
$\mathcal{O}(c\bar{n}(\bar{n}-1)/2)$, where $c$ is the number of cascades and $\bar{n}$ is the average cascade size. This search is too time-consuming as we analyze more on the experimental section. 

It should be noted that apart from pure IL, multiple methods have been developed to predict several aspects of a cascade, such as \textsc{Topo-LSTM} and \textsc{HiDAN} to predict the next infected node \cite{topo_rnn, wang2019hierarchical}, RMTPP, \textsc{DeepDiffuse} and CYANRNN for next node and time of infection \cite{rmtpp,deepdiffuse,cyanrnn}, \textsc{DeepCas} to predict cascade size \cite{deepcas}, and FOREST to predict next node and size together \cite{forest}.
These methods have advanced end-to-end neural architectures focused each on their respective tasks. However, their hidden representations can not be utilized in the context of IM, because they are derived by a sequence of nodes, hence we can not form individual influence relationships between two distinct nodes. In other words, an IM algorithm requires a node-to-node influence relationship, which is not clearly provided by the aforementioned references. Moreover, methods that do learn node-to-node influence but capitalize on other information such as sentiment of the context \cite{liu2019ct} can not be utilized as well due to our datasets missing content, but it is a promising future approach.

\begin{figure}[h!]
\centering
\includegraphics[width=0.45\textwidth]{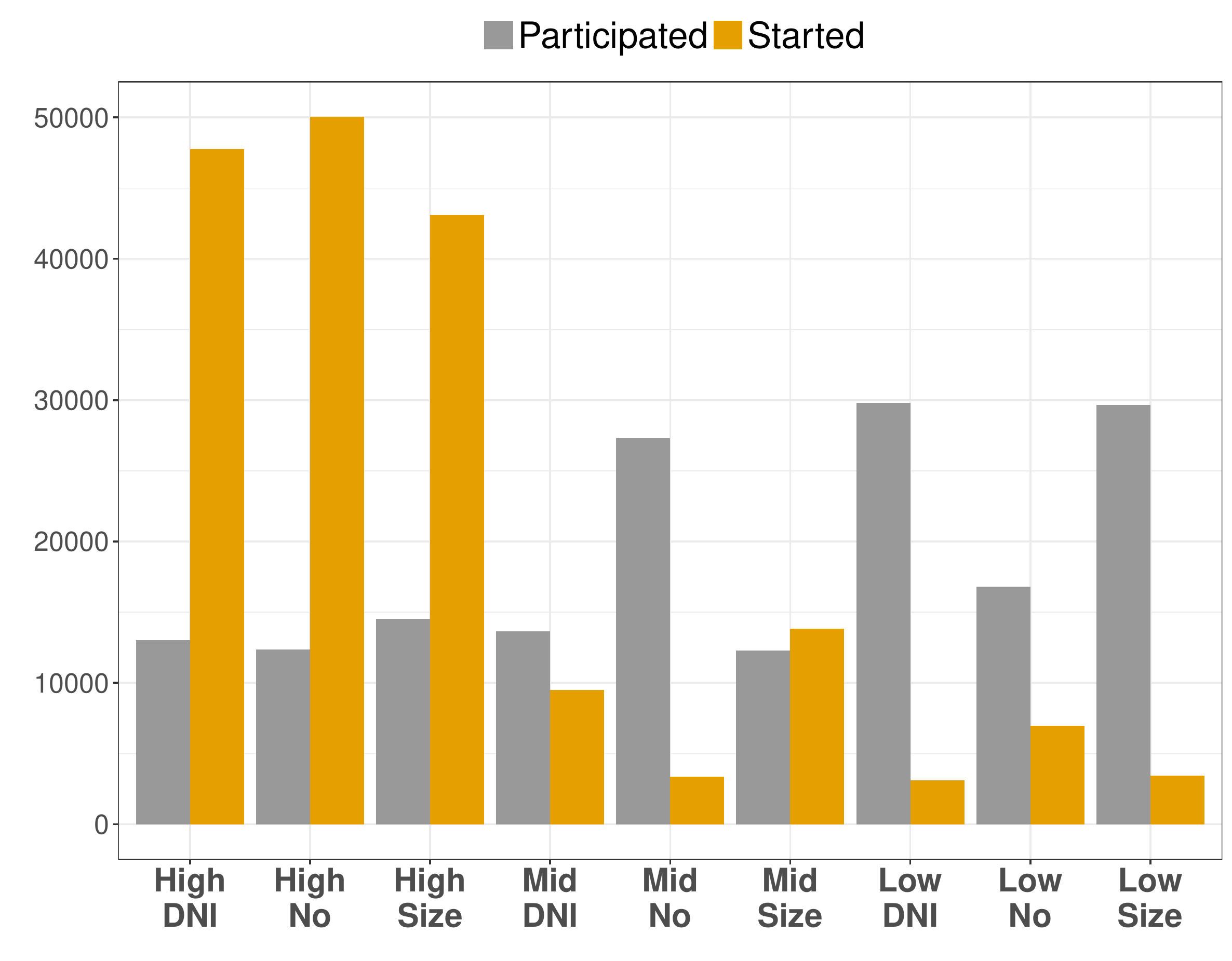}
\caption{Influencer's initiating versus participating in diffusion cascades. \textbf{DNI} stands for distinct nodes influenced, \textbf{size} stands for cascade size, and \textbf{no} stands for number of cascades started.}
\label{fig:initiator_barplot}%
\end{figure}

\vspace{1.3cm}
\IEEEraisesectionheading{
\section{Learning Influencer Vectors}
\label{sec:infector}}

\subsection{Node-Context Extraction}
Our goal in the first part of our methodology is to create a model that learns representations suitable for scalable model-independent IM.
Due to the lack of a diffusion model, the network's structure becomes secondary and online methods tend to rely only on the activity of individual nodes \cite{lagree2018algorithms,vaswani2017model}.
We follow suit and propose an IL method that focuses mainly on influencers.

\begin{figure*}[h]
\centering
\includegraphics[scale=.5]{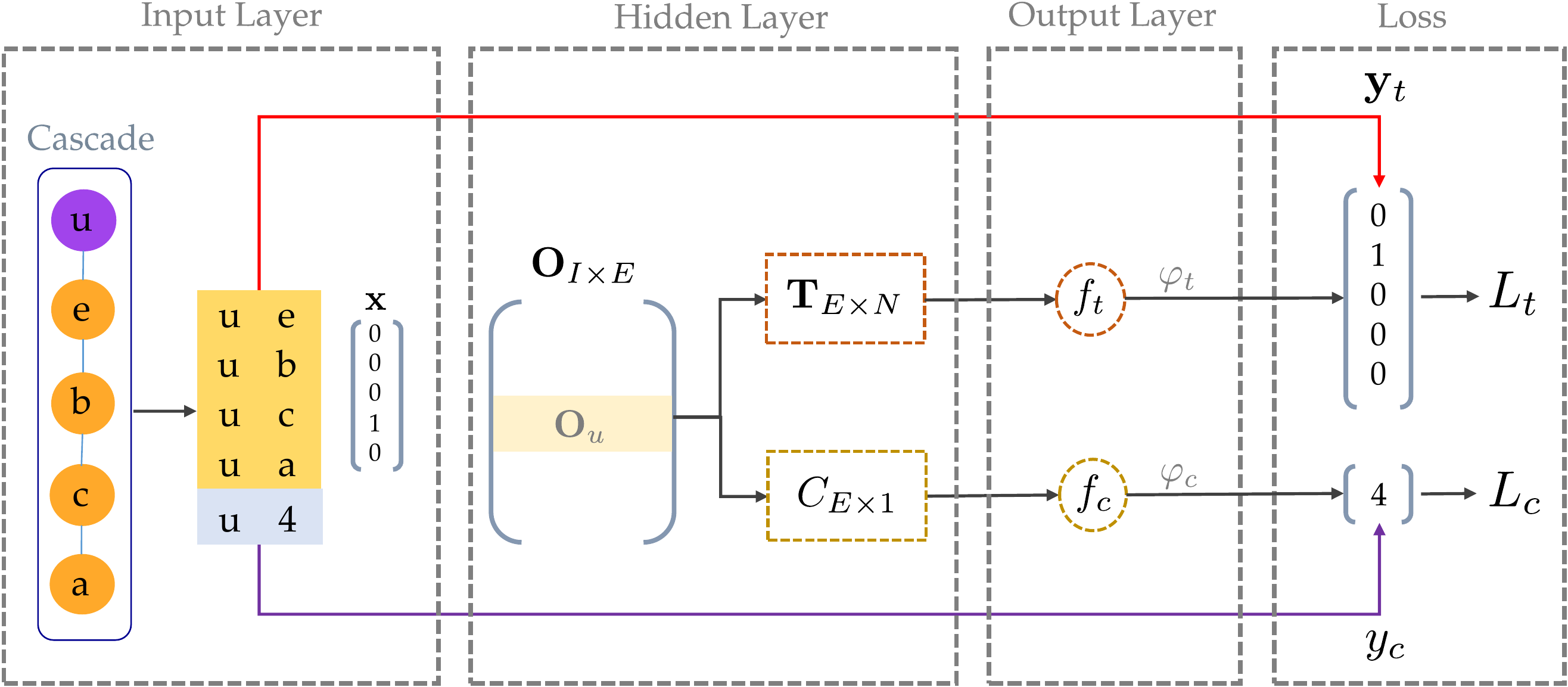}
\caption{Schematic representation of the INFECTOR model. Given the cascade, a sequence of pairs is extracted, each pair consisting of the initiator node $\mathbf{x}$ (one-hot embedding) which is the \textit{input}, and one of the "infected" nodes $\mathbf{y}_t$ i.e. e,b etc. which is the \textit{output}. The last pair is the initiator and the cascade size $y_c$ i.e. 4 as output. After the initator's embedding lookup through the origin embeddings $\mathbf{O}$, the vector passes through $\mathbf{T}$ and $f_t$ if the output is another node or through $\mathbf{C}$ and $f_c$ if the output is scalar. The loss functions are log-loss $L_t$ for node output or mean squared error $L_c$ for scalar output.}
 \label{fig:mtl_schema}
\end{figure*}

\begin{table}[t]
\centering
\begin{tabular}{p{1cm}p{5cm}p{1cm}}
    \toprule
      Symbol  & Meaning & Type\\
      \midrule
      $N$ & number of nodes & scalar\\
      $t_u$ & time of node' $u$ post & scalar\\
      $C_u$ & cascade initiated by $u$ & set\\
      $\mathbf{x}, \mathbf{y}_t$ & one-hot embedding of a node & vector\\
      $y_c$ & cascade length & scalar\\
      $X^t$ & $\mathbf{x},\mathbf{y}$ pairs extracted from cascade & list\\
      $\mathbf{O}$ & origin embeddings of INFECTOR & matrix \\
      $\mathbf{T}$ & target embeddings of INFECTOR & matrix \\
      $C$ & constant embeddings of INFECTOR & vector \\
      $p_{u,v}$ & probability of $u$ influencing $v$ & scalar \\
      $\mathbf{z}$ & hidden layer output & vector \\
      $f_t$ & softmax &  function \\
      $f_c$ & sigmoid &  function \\
      $\boldsymbol{\varphi}_t $ & output for node influenced &  vector \\
      $\varphi_c$ & output for cascade size & scalar\\
      $\mathbf{\Phi}$ & jacobian of INFECTOR's output & matrix\\
      $L$ &  loss of INFECTOR & vector \\
      $D$ & INFECTOR classification probabilities & matrix \\
      $\lambda_u$ & number of nodes to be influenced by $u$ &scalar \\
      $\hat{D_s}$ & $D$ sorted for the row of node $s$ & matrix \\
      $\sigma'(s)$ & influence spread of node set $s$ & scalar\\ 
      $S$ & Set of seed nodes & set \\
      \bottomrule
\end{tabular}
\caption{Table of symbols.}
\end{table}

We start from the context creation process that produces the input to the network.
In previous node-to-node IL models, initiating a cascade is considered equally important with participating in a cascade created by someone else \cite{feng2018inf2vec}, thus the context of a node is derived by the nodes occurring after it in a cascade.
This process requires the creation of the propagation network, meaning going through every node in the cascade and iterating over the subsequent nodes to search for a directed edge in the network. This has a complexity of $\mathcal{O}\Big(c(\bar{n}(\bar{n}-1)/2))$, where $c$ is the number of cascades and $\bar{n}$ is the average cascade size. 
Given that the average size of a cascade can surpass 60 nodes, it is a very time consuming for a scalable IM algorithm.
To overcome this we focus on the nature of IM, meaning the final chosen seed users will have to exert influence over other nodes, thus we may accelerate the process by capitalizing on the differences between influencers and simple users.
Intuitively, we expect the influencers to exhibit different characteristics in sharing content than the simple/susceptible nodes \cite{newman2002assortative}. 
We argue that an influencer's strength lies on the cascades she initiates, and evaluate this hypothesis through an exploratory analysis.
We utilize the cascades of \textit{Sina Weibo} dataset, a large scale social network accompanied by retweet cascades to validate our hypothesis. The dataset is split in train and test cascades based on their time of occurrence and each cascade represents a tweet and its set of retweets. We keep the 18,652 diffusion cascades from the last month of recording as a test set and the 97,034 from the previous 11 months as a train set. 
We rank all users that initiated a cascade in the test set based on three measures of success: the number of test cascades they spawn, their cumulative size, and the number of Distinct Nodes Influenced (DNI) \cite{continest,complex2018,panagopoulos2018diffugreedy}, which is the set of nodes that participated in these test cascades.
We bin the users into three categories based on their success in each metric, and for each category we compute the total cascades the users start in the training set opposed to those they participate in.

Here, we examine the contrast between the behavior of the influencers and normal users, meaning how more probable is for an influencer to start cascades compared to normal users. As we see in Figure \ref{fig:initiator_barplot}, users that belong to the top category of the test set are much more prone to create cascades compared to the ones who belong to the mid and low categories. Since our end goal is to find influencers for our algorithm, this observation allows us to focus solely on the initiators of the cascades, rather than every node in the cascade. Moreover, we observe that influencers in Weibo are more prone to create cascades opposed to participating in them. This means that by overlooking their appearances inside the cascades of others, we do not lose too much information regarding their influence relationships, as most of them start the cascades.


Thus, for our purpose, the context is extracted exclusively for the initiator of the cascade and will be comprised of all nodes that participate in it. 
Moreover, we will take into account the copying time between the initiator and the reposter, based on empirical observation on the effect of time in influence \cite{goyal2010learning}.
To be specific, all cascade initiators are called influencers, an influencer $u$'s context will be created by sampling over all nodes $v$ in a given cascade $c$ that $u$ started, with probability inversely proportional to their copying time. In this way, the faster $v$'s retweet is, the more probable it will appear in the context of $u$, following this formula $P(v|\mathcal{C}_u ) \sim\frac{(t_u-t_v)^{-1}}{\sum_{v' \in \mathcal{C}_u }{(t_u-t_v')^{-1}}}$.
It is a general principle to utilize temporal dynamics in information propagation when the network structure is not used \cite{netrate,cosine,continest} and is based on empirical observation on the effect of influence in time \cite{goyal2010learning}.
In our case we do an oversampling of 120\% to emphasize the importance of fast copying.
This node-context creation has a complexity of $\mathcal{O}(cn)$,  which is linear to the cascade's size and does not require searching in the underlying network.
Even more importantly, this type of context sets up the model to compute diffusion probabilities, i.e. influence between nodes with more than one hop distance in the network. 

\subsection{The INFECTOR Model}

The diffusion probabilities (DPs) are the basis for our model-independent approach, and exhibit important practical advantages over influence probabilities, as we analyze further below.
We use a multi-task neural network \cite{caruana1997multitask} to learn simultaneously the aptitude of an influencer to create long cascades as well as the diffusion probabilities between her and the reposters. We chose to extend the typical IL architecture in a multi-task learning setting because ($i$) the problem could naturally be broken in two tasks, and ($ii$) theoretical and applied literature suggests that training linked tasks together improves overall learning \cite{evgeniou2004regularized,panagopoulos2017multi}.
In our case, given an input node $u$, the first task is to classify the nodes it will influence and the second to predict the size of the cascade it will create.
An overview of our proposed INFECTOR model can be seen in Figure \ref{fig:mtl_schema}.
There are two types of inputs. The first is the training set comprised of the node-context pairs. As defined in the previous section, given a cascade $t$ with length $m$, we get a set  $\mathbf{X}^t=\{(\mathbf{x}^1,\mathbf{y}^1_t),(\mathbf{x}^1,\mathbf{y}^2_t),...,(\mathbf{x}^m,\mathbf{y}^m_t)\}$, where $\mathbf{x}\in \mathbb{R}^I$ and  $\mathbf{y}_t \in \mathbb{R}^N$ are one hot encoded nodes, with $I$ the number of influencers in the train set and $N$ the number of nodes in the network. The second is a similar set $X^c$, where instead of a vector e.g. $\mathbf{y}^1_t$, there is a number $\mathbf{y}^1_c$ denoting the length of that cascade, initiated by the $\mathbf{x}^1$. To perform joint learning of both tasks we mix the inputs following the natural order of the data; given a cascade, we first input the influencers-context pairs extracted from it and then the influencers-cascade length pair, as shown in Figure \ref{fig:mtl_schema}.
$\mathbf{O} \in \mathbb{R}^{I \times E}$, with $E$ being the embeddings size, represents the source embeddings, $\mathbf{O}_u \in \mathbb{R}^{1 \times E}$ the embeddings of cascade initiator $u$, $\mathbf{T} \in \mathbb{R}^{E \times N}$ the target embeddings and $C \in \mathbb{R}^{E \times 1} $ is a constant vector initialized to 1. Note that $\mathbf{O}_u$ is retrieved by the multiplication of the one-hot vector of $u$ with the embedding matrix $\mathbf{O}$.
The first output of the model represents the diffusion probability $p_{u,v}$ of the source node $u$ for a node $v$ in the network. It is created through a softmax function and its loss function is the cross-entropy. The second output aims to regress the cascade length, which has undergone min-max normalization relative to the rest of the cascades in this set, and hence a sigmoid function is used to bound the output at $(0,1)$. The respective equations can be seen in Table \ref{tab:neurons}.
Here $y_t$ is a one-hot representation of the target node and $y_c$ is the normalized cascade length.
We employ a non-linearity instead of a simple regression because without it, the updates induced to the hidden layer from the second output would heavily overshadow the ones from the first. Furthermore, we empirically observed that the update of a simple regression would cause the gradient to explode eventually. 
Moreover, variable $C$ is a constant vector of ones that is untrainable, in order for the update of the regression loss to change only the hidden layer. This change was motivated by experimental evidence.

\begin{table}[h!]
\centering
\begin{tabular}{p{1cm}p{3.5cm}p{3cm}}
    \toprule
     & \textit{Classify Influenced Node}  & \textit{Regress Cascade Size}\\
      \midrule
      \textbf{Hidden}& $\mathbf{z}_{t,u} = \mathbf{O}_u\mathbf{T}+b_t$ & $\mathbf{z}_{c,u} = \mathbf{O}_uC+b_c$ \\
      \midrule
      \textbf{Output}	& $\boldsymbol{\varphi}_t = \frac{e^{(\mathbf{z}_{t,u})}}{\sum_{u' \in G}{e^{\mathbf{z}_{t,u'}}}}$ & 
      $\boldsymbol{\varphi}_c= \frac{1}{1+e^{-(\mathbf{z}_{c,u})}}$ \\
      \midrule
      \textbf{Loss} & $L_t = \mathbf{y}_t \log(\boldsymbol{\varphi}_t)$ & 
      $L_c = ({y_c-\varphi}_c)^2$\\ 
      \bottomrule
\end{tabular}

\caption{The layers of INFECTOR.	\label{tab:neurons}}
\end{table}

The training happens in an alternating manner, meaning when one output is activated the other is idle, thus only one of them can change the shared hidden layer at a training step.
For example, given a node $u$ that starts a cascade of length 4 like in Figure \ref{fig:mtl_schema}, $\mathbf{O}_u$ will be updated 4 times based on the error of $\frac{\partial L_t}{\partial \mathbf{O}}$, and one based on $\frac{\partial L_c}{\partial \mathbf{O}}$, using the same learning rate $e$  for the training step $\tau$:
\begin{equation}
\mathbf{O}_{u,\tau+1} =\mathbf{O}_{u,\tau} - e\frac{\partial L}{\mathbf{\partial O}} 
\end{equation}

\noindent The derivatives from the chain rule are defined as:
\begin{equation}
\frac{\partial L}{\partial \mathbf{O}} = \frac{\partial L}{\partial \varphi}\frac{\partial \varphi}{\partial z}\frac{\partial z}{\partial \mathbf{O}}.
\label{eq:dl_ds}
\end{equation}

\noindent The derivatives in Eq. (\ref{eq:dl_ds}) differ between regression and classification. For the loss function of classification $L_c$, we have that 
\begin{equation}
\frac{\partial \mathbf{z}_c}{\partial \mathbf{O}}=\textbf{T}^\top
\label{eq:dzc}
\end{equation}

\begin{equation}
\begin{split}
\frac{\partial \boldsymbol{\varphi}_t}{\partial \mathbf{z}_t}=
\begin{cases} 
\boldsymbol{\varphi}_t^{u} (1-\boldsymbol{\varphi}_t^v), \text{~~~~~$u=v$} \\
-\boldsymbol{\varphi}_t^{u} \boldsymbol{\varphi}_t^v, \text{~~~~~~~~~~~$u\neq v$}
\end{cases}
\end{split}
\end{equation}

\noindent where $u$ is the input node and $v$ is each of all the other nodes, represented in different dimensions of the vectors $\boldsymbol{\varphi}_t$. If we create a matrix consisting of $N$ replicates of $\boldsymbol{\varphi}_t$, we can express the Jacobian as this dot product: 
\begin{equation}
\frac{\partial \boldsymbol{\varphi}_t}{\partial \mathbf{z}_t} = {\mathbf{\Phi}_t}\cdot(I-{\mathbf{\Phi}_t})^\top, ~~~  \mathbf{\Phi}_t = \begin{bmatrix}\boldsymbol{\varphi}_t\\ \vdots\\ \boldsymbol{\varphi}_t\end{bmatrix}  \in  \mathbb{R}^{N \times N}
\label{eq:dft}
\end{equation}

\begin{equation}
 \frac{\partial L_t}{\partial \boldsymbol{\varphi}_t} = \mathbf{y_t} \Bigg(-\frac{1  }{\mathbf{\boldsymbol{\varphi}_t}}\Bigg) 
\label{eq:dlt}
\end{equation}

\noindent The derivatives for the regression task are:
\begin{equation}
    \frac{\partial \mathbf{z}_c}{\partial \mathbf{O}}=C^\top
    \label{eq:dzt}
\end{equation}

\begin{align}
\begin{split}
\frac{\partial \varphi_c }{\partial \mathbf{z}_c} = 
\frac{1}{1+e^{-\mathbf{z}_{c}}}\Bigg( 1-\frac{1}{1+e^{-\mathbf{z}_{c}} }\Bigg)=
\varphi_c(1-\varphi_c)
\end{split}
\label{eq:dfc}
\end{align}

\begin{equation}
\frac{\partial L_c}{\partial \varphi_c} = 2(y_c-\varphi_c) \frac{(-\varphi_c)}{\partial \varphi_c}=-2(y_c-\varphi_c)
\label{eq:dlc}
\end{equation}

\noindent From Eqs. (\ref{eq:dl_ds}), (\ref{eq:dzc}), (\ref{eq:dft}), (\ref{eq:dlt}), (\ref{eq:dzt}), (\ref{eq:dfc}),(\ref{eq:dlc}) we have:

\begin{equation}
\frac{\partial L}{\partial \mathbf{O}} =
\begin{cases}
    -\frac{ y_t }{\boldsymbol{\varphi}_t } ({\mathbf{\Phi}_t }\cdot(I-{\mathbf{\Phi}_t})^\top) \mathbf{T}^\top\\
    -2(y_c-\varphi_c)(\varphi_c(1-\varphi_c))\mathbf{C}^\top
    \end{cases}
    \label{eq:gradients}
\end{equation}

As mentioned above, the aim here is for the embedding of an influencer (hidden layer $\mathbf{O}$) to not only capture who she influences but also her overall aptitude to create lengthy cascades. To be specific, for each node $v$ inside the cascade, \textbf{$O_u$}  and \textbf{$T_v$}  will undergo updates using the upper formula from Eq. \eqref{eq:gradients} to form the diffusion probabilities, depending on $\boldsymbol{\varphi_t}$ and $\boldsymbol{y_t}$. We see that the upper update is formed based on a combination of the gradients, hence each dimension of the update vector might be positive or negative. In contrast, when updating the parameters for the regression task, we multiply a constant vector $C$ with a scalar value. Thus, the update will increase or decrease the overall norm of the embedding analogously to the difference of the predicted and the actual cascade size.

\begin{figure}
\centering
\includegraphics[width=.2\textwidth]{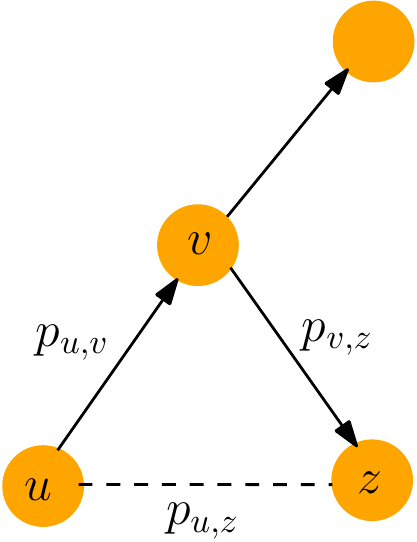}
\caption{Example of higher order influence showing the difference between influence and diffusion probabilities. \label{fig:toy_influence}}
\end{figure}

The main difference between INFECTOR and similar node-to-node IL methods is that it computes diffusion probabilities and does not require the underlying social network, in contrast to influence probabilities which are assigned to edges of the network \cite{bourigault2016representation,feng2018inf2vec,saito2009learning}.
 Intuitively, DP is the probability of the susceptible node appearing in a diffusion started by the influencer, independently of the two nodes' distance in the network. This means that the underlying influence paths from the seed to the infected node is included \textit{implicitly}, which changes drastically the computation of influence spread.
For example, in a setting with diffusion models and influence probabilities, a node $u$ might be able to influence another node $z$ by influencing node $v$ between them, as shown in Figure \ref{fig:toy_influence}. However, in our case, if $u$ could indeed induce the infection of $z$ in a direct or indirect manner, it would be depicted by the diffusion probability $p_{u,z}$.

This approach captures the case when $v$ appears in the cascades of $u$ and $z$ appears in the cascades of $v$ but not in $u$'s. This is a realistic scenario that occurs when $v$ reshares various types of content, in which case content from $u$ might diffuse in different directions than $z$. Hence, it would be wrong to assume that $u$'s infection would be able to eventually cause $z$'s. Typical IM algorithms fail to capture such higher order correlations because diffusion models act in a Markovian manner and can spread the infection from $u$ to $z$.
Apart from capturing higher order correlations, DPs will allow us to overcome the aforementioned problems induced in IM by the diffusion models as well as surpass the computational bottleneck of the repeated simulations. 
Please note that our method's final purpose is influence maximization, which is why we focus so much on the activity of influencers and overlook the rest of the nodes. If we aimed for a purely prediction task, such as cascade size or next infected node prediction, it is not clear that our learning mechanism would be effective.

\vspace{1.5cm}
\IEEEraisesectionheading{
\section{Influence Maximization with Influencer Vectors} 
\label{sec:im}}

\subsection{Diffusion With Influencer Vectors}

In the second part of the methodology, we aim to perform fast and accurate IM using the representations learnt by \textsc{INFECTOR} and their properties.
Initially, we will use the combination of the representations which form the \textit{diffusion probability} of every directed pair of nodes.
The DPs derived from the network can define a \textit{matrix}:
\begin{equation}
\mathbf{D} = \begin{bmatrix}f_t(\mathbf{O}_1 T)\\\vdots\\ f_t(\mathbf{O}_I T)\end{bmatrix}
 \label{eq:d_matrix}
\end{equation}
which consists of the nodes that initiate train cascades (influencers) in one dimension and all susceptible nodes in the other. Even though influencers are fewer than the total nodes, $\mathbf{D}$ can still be too memory-demanding for real-world datasets. To overcome this, we can keep the top $P\%$ influencers based on the norm of their influencer embedding $|\mathbf{O}_u|$ to reduce space, depending on the device. Recall that, the embeddings are trained such that their norm captures the influencers' potential to create lengthy cascades, because $\mathbf{O}_uC=\sum_{i}^E\mathbf{O}_u(i) =|O|$, since $C$ is constant. 

Subsequently, $\mathbf{D}$ can be interpreted as a bipartite network where the left side nodes are the candidate seeds for IM, and each of them can influence every node in the right side, where the rest of the network resides. Since all edges are directed from left to right, no paths with length more than 1 exist. This means that the probability of an edge can only define the infection of one node, and it is independent of the infection of the rest---hence, we do not require a diffusion model to estimate the spread. 

To be specific regarding our assumptions and the difference with diffusion models, we clarify that typically simulations are performed in order to account for all $2^e$ possible combinations of edges. This happens because an edge might produce the infection of more than the target node e.g. when it is a bridge between two communities. This can not happen in our setting by the definition of diffusion probability: if a node can influence another node, no matter how far that is in the original network, it will be depicted in their edge weight. That is clear in the first iteration, where the infection of a seed can only cause the infection of its neighbors hence simulations are not required. To keep this property in the subsequent iterations (where the candidate's influenced spread overlaps with previous seeds) we make the assumption that each node will influence for sure a certain number of nodes, based on INFECTOR's representations. Hence these nodes can be removed, as they should not be recalculated in the influence spread of subsequent candidate seeds.

To formulate that influence spread, we will utilize the embedding of candidate seed $u$'s to compute the fraction of all $N$ nodes it is expected to influence based on its $|\mathbf{O}_u|$:
\begin{equation}
    \lambda_u =\Bigg\lceil N \frac{||\mathbf{O}_u||_2}{\sum_{u' \in \mathcal{I}} ||\mathbf{O}_{u'}||_2} \Bigg \rceil
    \label{eq:exp_spread}
\end{equation}
where $\mathcal{I}$ is the set of candidate seeds. The term resembles the norm of $u$ relative  to the rest of the influencers and the size of the network. It is basically computing the amount of the network $u$ will influence.
Since we know a seed $s$ can influence a certain number of nodes, we can use the diffusion probabilities to identify the top $\lambda_s$ nodes it connects to. 
Moreover, we have to take into account the values of the DPs, to refrain from selecting nodes with big $\lambda_s$ but overall small probability of influencing nodes.
Consequently, the influence spread is defined by the sum of the top $\lambda_s$ DPs:
\begin{equation}
    \sigma'(s) = \sum_j^{\lambda_s}{\mathbf{\hat{D}}_{s,j}}
    \label{eq:new_sigma}
\end{equation}
where $\mathbf{\hat{D}}_s$ is the DPs of seed $s$ sorted in descending order. Once added to the seed set, as mentioned above the seed's influence set is considered infected and is removed from $\mathbf{D}$.
This means that a seed's spread will never get bigger in two subsequent rounds, and we can employ the CELF trick \cite{celf} to accelerate our solution. 
It should be noted that one can either model alternatively the influence spread based on the total probability of each node getting influenced, which results in a non-convex function, or approach the problem as weighted bipartite matching. Both these cases are analyzed briefly in the Appendix \ref{sec:appendix}

To retain the theoretical guarantees of the greedy algorithm $(1-1/e)$, we need to prove the monotonicity and submodularity of the influence spread that we optimize in each iteration. To do this, we need to separate the influence spread of the seed set $S$ throughout the different steps of the algorithm.
Let $i(s)$ be the step before node $s$ was inserted in $S$, $S^i$ be seed set at step $i$, $\mathbf{D^{i}}$ the DP matrix at step $i$ and $u$ the current candidate seed. 

\begin{corollary}
The influence spread is monotonic increasing.
\end{corollary}
\begin{IEEEproof}
\begin{align}
\begin{split}
 \sigma'(S^{i(u)}\cup \{u\}) = \sum_{s\in S^{i(u)}\cup \{u\}}\sum_j^{\lambda_s}{\mathbf{\hat{D}}^{i(s)}_{s,j}} = \\
 \sum_{s\in S^{i(u)}}\sum_j^{\lambda_s}{\mathbf{\hat{D}}^{i(s)}}_{s,j}+
 \sum_j^{\lambda_u}{\mathbf{\hat{D}}^{i(u)}_{u,j}} =\\
 \sigma'(S^{i(u)}) + \sum_j^{\lambda_u}{\mathbf{\hat{D}}^{i(u)}_{u,j}} \geq \sigma'(S^{i(u)})
\end{split}
\end{align}
\end{IEEEproof}

\begin{corollary}
The influence spread is submodular.
\end{corollary}
\begin{IEEEproof}
\begin{align}
\begin{split}
\sigma'(S^{i(u)-1}\cup \{u\}) - \sigma'(S^{i(u)-1})  = \\
\sum_{s\in S^{i(u)-1} \cup \{u\}}\sum_j^{\lambda_s}{\mathbf{\hat{D}}^{i(s)}_{s,j}} - \sum_{s\in S^{i(u)-1}}\sum_j^{\lambda_s}{\mathbf{\hat{D}}^{i(s)}_{s,j}} =  \sum_j^{\lambda_u}{\mathbf{\hat{D}}^{i(u)-1}_{u,j}}
\end{split}
\label{eq:sub1}
\end{align}
\begin{equation}
    \sigma'(S^{i(u)} \cup \{u\}) - \sigma'(S^{i(u)})  =\sum_j^{\lambda_u}{\mathbf{\hat{D}}^{i(u)}_{u,j}}
    \label{eq:sub2}
\end{equation}
\noindent $\mathbf{\hat{D}}^{i(u)}_{u,j}, \forall j \in \lambda_u$ are the top DPs of $u$ for nodes that are left uninfected from the seeds in $S^{i(u)}$. 
Note here that, $\mathbf{\hat{D}}^{i(u)}_{u,j}, \forall j \in \lambda_u$ are the top DPs of $u$ for nodes that are left uninfected from the seeds in $S^{i(u)}$. So, the influence spread of $u$ is reverse analogous to the influence spread of the seed set up to that step:
\begin{align}
\begin{split}
S^{i(u)-1}\subseteq S^{i(u)}  \Leftrightarrow \sum_{s\in  S^{i(u)-1}}\sum_j^{\lambda_s}{\mathbf{\hat{D}}^{i(s)}_{s,j}} \leq \sum_{s\in  S^{i(u)}}\sum_j^{\lambda_s}{\mathbf{\hat{D}}^{i(s)}_{s,j}} \Leftrightarrow\\
\sum_j^{\lambda_u}{\mathbf{\hat{D}}^{i(u)-1}_{u,j}} \geq \sum_j^{\lambda_u}{\mathbf{\hat{D}}^{i(u)}_{u,j}}
\end{split}
\label{eq:sub}
\end{align}
\end{IEEEproof}

\subsection{IMINFECTOR}
The algorithm is given at Algorithm \ref{alg:iminfector} and is an adaptation of CELF using the proposed influence spread and updates on $\mathbf{D}$. We keep a queue with the candidate seeds and their attributes (line 2) and the uninfected nodes 
(line 3). The attributes of a candidate seed are its influence set (line 5) and influence spread (line 6), as defined by eq. (\ref{eq:new_sigma}). We sort based on the influence spread (line 8) and proceed to include in the seed set the candidate seed with the maximum marginal gain, which is $\Omega$ after the removal of the infected nodes. Once a seed is chosen, the nodes it influences are marked as infected (line 14). The DPs of the candidate seeds are reordered after the removal of the infected nodes (line 17)  and the new marginal gain is computed (line 18).

An alternative to Eq. (\ref{eq:exp_spread}) would be to keep the actual percentage of nodes influenced by each seed, i.e., $||\mathbf{O}_u||_2/N$. This approach, however, suffers from a practical problem. In real social networks, influencers can create cascades with tens or even hundreds of thousands of nodes. In this case, $|\lambda_u|$ can be huge, so when our algorithm is asked to come up with a seed set of e.g., 100 nodes, matrix $\mathbf{D}$ will be emptied in the first few iterations that will include seeds covering the whole network. This would constrain our algorithm to small seed set sizes, in which case simple ranking metrics tend to be of equal strength with IM,
as we argue further on the evaluation methodology. Hence, we use a normalization that alleviates moderately the large differences between influence spreads and allows for bigger seed set sizes.

\begin{algorithm}[h!]
\caption{ \textsc{IMINFECTOR}}
\label{alg:iminfector}
\textbf{Input:} Probability diffusion matrix $D$, expected influence spread $\lambda$, seed set size $\ell$\\
\textbf{Output:} Seed set $\mathcal{S}$

\begin{algorithmic}[2]
\Procedure{\textsc{IMINFECTOR}}{$D,\lambda,\ell$}
\State set $q \gets [~],\mathcal{S} \gets \emptyset $
\State $F = 0:\text{dim}(D)[1]$
\For{$s = 0;s<\text{dim}(D)[0];s++$} 
\State $R \gets \text{argsort}(D[s,F])[0:\lambda[s]]$
\State $\Omega \gets \text{sum}(D[s,R])$
\State $q.\text{append}([s,\Omega,R,0])$
\EndFor
\State $q \gets \text{sort}(q,1)$
\While{$|\mathcal{S}|<\ell$}
\State $u \gets q[0]$
 \If{$u[3]== |\mathcal{S}|$}
 \State $\mathcal{S}.\text{add}(u[0])$
 \State $R \gets u[2]$ 
 \State $F \gets F- R$
 \State $q.\text{delete}(u)$
\Else
\State $R \gets\text{argsort}(D[u[0],F])[0:\lambda[u[0]]]$
\State $\Omega \gets \text{sum}(D[u[0],R])$
\State $q[0] \gets [u[0],\Omega,R,|\mathcal{U}|]$
\State $q \gets \text{sort}(q,1)$
\EndIf
\EndWhile
\State \textbf{return} $\mathcal{S}$
\EndProcedure
\end{algorithmic}
\end{algorithm}

\tikzset{axis/.style={Black!75!black, -latex, shorten <=0cm, shorten >=-2*\nudge cm}}
\tikzset{line/.style={thick,green}}

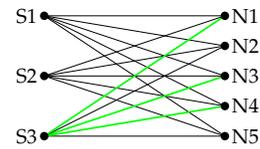
\begin{figure}[h!]
    \vspace{.3cm}
    \resizebox{0.9\columnwidth}{!}{
    \begin{tikzpicture}
    \tikzstyle{cell} = [rectangle,draw,baseline]
    \matrix[matrix,
    matrix of math nodes,
    first column text width/.code={%
      \tikzset{
        column 1/.style={
          nodes={text width=3em}
        },
      }
    },
    first column text width=4em,
    nodes={cell,minimum height=1.5em, text width=3.5em,
                anchor=center, align=center, 
                inner sep=0pt, outer sep=0pt},
    row sep =-\pgflinewidth,
    column sep = -\pgflinewidth,
    ampersand replacement=\&
  ] at (-1.4,2.5) (am)
  {
  \text{Seed} \& N1 \& N2 \& N3 \& N4 \& N5 \& \lambda \&\sigma'\\
  ~{\bf S1}~ \& ~0.1 \& ~0.3 \& ~0.2\& ~0.2\& ~0.2\& 2 \&0.5 \\
 ~{\bf S2 }~ \& ~0.4 \& ~0.2 \& ~0.2\& ~0.1\& ~0.2\& 2\&0.6 \\
 ~{\bf S3}~ \& ~\textcolor{blue}{0.5} \& ~0.1 \& ~\textcolor{blue}{0.2}\& ~\textcolor{blue}{0.2}\& ~0\& 3 \& \textcolor{blue}{0.9}\\};

\vspace{.2cm}
\hspace{3.5cm}

    \draw (-6,1) -- (-3,-1) coordinate (ineq1); 
    \draw (-6,1) -- (-3,-0.5) coordinate (ineq1);
    \draw (-6,1) -- (-3,-0) coordinate (ineq1);
    \draw (-6,1) -- (-3,0.5) coordinate (ineq1);
    \draw (-6,1) -- (-3,1) coordinate (ineq1);
    
    \draw (-6,0) -- (-3,-1) coordinate (ineq1);
    \draw (-6,0) -- (-3,-0.5) coordinate (ineq1);
    \draw (-6,0) -- (-3,-0) coordinate (ineq1);
    \draw (-6,0) -- (-3,0.5) coordinate (ineq1);
    \draw (-6,0) -- (-3,1) coordinate (ineq1);
    
    \draw (-6,-1) -- (-3,-1) coordinate (ineq1);
    \draw[line] (-6,-1) -- (-3,-0.5) coordinate (ineq1);
    \draw[line] (-6,-1) -- (-3,-0) coordinate (ineq1);
    \draw (-6,-1) -- (-3,0.5) coordinate (ineq1);
    \draw[line] (-6,-1) -- (-3,1) coordinate (ineq1);
    
    \foreach \coord/\adj/\name in {
      {( -6,1 )}/left/S1, 
      {(-6,0)}/left/S2, 
      {(-6,-1)}/left/S3, 
      {(-3,-1)}/right/N5,
      {(-3,-0.5)}/right/N4,
      {(-3,-0)}/right/N3,
      {(-3,0.5)}/right/N2,
      {(-3,1)}/right/N1
    } {
      \fill \coord circle (2pt) node[\adj] {\name};
    };
    \end{tikzpicture}
    }
    \caption{Example of \textsc{IMINFECTOR}: step 1.}
    \label{fig:spread1}
\end{figure}

To give a concrete example of how the algorithm works, let's imagine a simple network with 8 nodes, as depicted in Figure \ref{fig:spread1}, with the respective weights from each candidate seed $S$ to each susceptible node $N$. In the first step, $\lambda$ of a seed $S$ defines how many of the $S$'s top susceptible nodes should be taken into account in the computation of $\sigma'$. In this case, we keep the top 3 for $S3$ and top 2 for $S2$ and $S1$. The top 3 for $S3$ correspond to $\{N1,N3,N4\}$,  as indicated by the green edges in the bipartite network, whose sum gives a $\sigma'=0.9$. For $S2$, we could  either take $\{N1,N2\}$ or $\{N1,N3\}$, both giving a $\sigma'=0.6$, which is lower than $S3$, and the same applies for $S1$. Thus $S3$ is chosen as the seed for step 1 and the influenced nodes are removed.


\begin{figure}[h!]
      \resizebox{0.9\columnwidth}{!}{
    \begin{tikzpicture}
    \tikzstyle{cell} = [rectangle,draw,baseline]
    \hspace{-1cm} 
    \matrix[matrix,
    matrix of math nodes,
    first column text width/.code={%
      \tikzset{
        column 1/.style={
          nodes={text width=3em}
        },
      }
    },
    first column text width=4em,
    nodes={cell,minimum height=1.5em, text width=3.5em,
                anchor=center, align=center, 
                inner sep=0pt, outer sep=0pt},
    row sep =-\pgflinewidth,
    column sep = -\pgflinewidth,
    ampersand replacement=\&
  ] at (-1.2, 2.5) (am)
  {
  \text{Seed} \& N2 \& N5 \& \lambda \& \sigma'\\
  ~{\bf S1} \& ~\textcolor{blue}{0.3} \& ~\textcolor{blue}{0.2} \& ~2 \&~ 0.5\\ 
 ~{\bf S2}  \& ~0.2  \& ~0.2 \&~ 2 \& ~0.4 \\};

\hspace{4cm} 
    
    \draw (-6,0) -- (-4,0) coordinate (ineq1);
    \draw (-6,0) -- (-4,1) coordinate (ineq1);
    \draw[line] (-6,1) -- (-4,0) coordinate (ineq1);
    \draw[line] (-6,1) -- (-4,1) coordinate (ineq1);
    
   \foreach \coord/\adj/\name in {
      {(-6,1)}/left/S1,
      {(-6,0)}/left/S2,
      {(-4,0)}/right/N5,
      {(-4,1)}/right/N2
    }{
      \fill \coord circle (2pt) node[\adj] {\name};
    };
    
    \end{tikzpicture}
    }
    \caption{Example of \textsc{IMINFECTOR}: step 2.}
    \label{fig:spread2}
\end{figure}
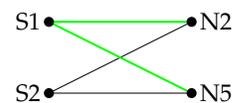

In the next step, matrix $\mathbf{D}$ is updated to remove the influenced nodes and $\sigma'$ is recomputed. In this case, there are only two nodes left for each seed, so their influence spread is computed by the sum of all their edges, since their $\lambda=2$. An important point here is that while in the first step $S2$ had a stronger $\sigma'$ than $S1$, $S1$ is stronger now, due to the removal of $N1$, which was strongly susceptible to the influence of $S2$, but infected in the previous round by $S3$. Thus, the chosen seed in step 2 is $S1$.


\noindent{\textbf{Running time complexity.}} The first step of the algorithm is to compute the influence spread for all candidate seeds $I$ in $\mathbf{D}$, which takes $I \cdot N\log N$, because we sort the DPs of each candidate seed to take the top $\lambda$. The sorting of $q$ which has $\mathcal{O}(I \cdot \log I)$ before the algorithm's iterations start, adds constant time to the total complexity.
With that in mind, and taking into acount the seed set size $\ell$, the complexity is analogous to $\mathcal{O}(\ell \cdot I(I \log I) \cdot (N \log N) )$, similar to CELF, but with sorting $N$ in every evaluation of the influence spread. In practice, in every iteration, $N$ is diminished by an average of $\lambda$ because of the removed influence set, so the final logarithmic term is much smaller.
Finally, due to the nature of CELF, much fewer influence spread evaluations than $I$ take place in reality (line 16 in the algorithm), which is why our algorithm is fast in practice.

\vspace{1.3cm}
\IEEEraisesectionheading{
\section{Experimental Evaluation} 
\label{sec:exp}}

\subsection{Datasets}
It is obvious that we can not use networks from the IM literature because we require the existence of ground truth diffusion cascades. Although such open datasets are still quite rare, we managed to assemble three, two social and one bibliographical, to provide an estimate of performance in different sizes and information types. Table \ref{tab:fields} gives an overview of the datasets.  

\begin{table}[h!]
\centering
\begin{tabular}{p{2.2cm}p{1.6cm}p{1.6cm}p{1.6cm}}
    \toprule
     & \textit{Digg}  & \textit{MAG} &\textit{Sina Weibo}\\
      \midrule
      \textbf{Nodes}& 279,631 & 1,436,158 & 1,170,689 \\
      \midrule
      \textbf{Edges} & 2,251,166 & 15,928,078 & 225,877,808 \\
      \midrule
      \textbf{Cascades} & 3,553 & 181,020 & 115,686 \\
       \midrule
       \textbf{Avg Cascade size} & 847 & 29 & 148 \\
      \bottomrule
\end{tabular}

\caption{Summary of the datasets used.\label{tab:fields}}
\end{table}

\begin{itemize}
    \item \textit{Digg}: 
    A directed network of a social media where users follow each other and a vote to a post allows followers to see the post, thus it is treated as a retweet \cite{digg}. 
    \item \textit{MAG Computer Science}: 
    We follow suit from \cite{deepinf} and define a network of authors with undirected co-authorship edges, where a cascade happens when an author writes a paper and other authors cite it. In other words, a co-authorship is perceived as a friendship, a paper as a post and a citation as a repost. In this case, we the employ Microsoft Academic Graph \cite{sinha2015overview} and filter to keep only papers that belong to computer science.  We remove cascades with length less than 10. 
     \item \textit{Sina Weibo}: 
        A directed follower network where a cascade is defined by the first tweet and the list of retweets \cite{weibo}. We remove from the network nodes that do not appear in the cascades i.e. they are practically  inactive, in order to make more fair the comparison between structural and diffusion-based methods, since the evaluation relies on unseen diffusions. 
\end{itemize}


\subsection{Baseline Methods}
Most traditional IM algorithms, such as greedy \cite{kempe2003maximizing} and CELF \cite{celf}, do not scale to the networks we have used for evaluation. Thus, we employ only the most scalable approaches:

\begin{itemize}
    \item \textsc{K-Cores}: The top nodes in terms of their core number, as defined by the undirected $k$-core decomposition. This metric is extensively used for influencer identification and it is considered the most effective structural metric for this task \cite{malliaros2016locating, core-vldb20}. 
    \item \textsc{AVG Cascade Size}: The top nodes based on the average size of their cascades in the train set. This is a straightforward ranking of the nodes that have proven effective in the past \cite{bakshy2011everyone}.     
    \item \textsc{SimPath}: A heuristic that capitalizes on the locality of influence pathways to reduce the cost of simulations in the influence spread \cite{simpath}. (The \textit{threshold} is set to 0.01).
    \item \textsc{IMM}: An algorithm that derives reverse reachable sets to accelerate the computation of influence spread while retaining theoretical guarantees \cite{imm}. It is considered state-of-the-art in efficiency and effectiveness. (Parameter $\epsilon$  is set to  0.1). 
     \item \textsc{Credit Distribution}: 
    This model uses cascade logs \textit{and} the edges of the network to assign influence credits and derive a seed set \cite{credit}. (Parameter $\lambda$  is set to 0.001).
    \item \textsc{CELFIE}: The previous version of the proposed algorithm, which relies on influence learning and sampling-based influence spreading \cite{panagopoulos2020influence}.
    \item \textsc{IMINFECTOR}: Our proposed model based solely on cascades, with embedding size equal to 50, trained for 5 epochs with a learning rate of 0.1. The reduction percentage $P$ is set to 10 for \textit{Weibo} and \textit{MAG}, and 40 for \textit{Digg}.
\end{itemize}
The parameters are the same as in the original papers.
 Comparing network-based methods such as \textsc{SimPath} and \textsc{IMM} with methods that use both the network and cascades, is not a fair comparison as the latter capitalize on more information. To make this equitable, each network-based method is coupled with two influence learning (IL) methods that provide the IM methods with influence weights on the edges, using the diffusion cascades:
 
\begin{itemize}
        \item \textsc{Data-based (DB)}:  Given that the follow edge $u\rightarrow v$ exists in the network, the edge probability is set to
        $A_{u \rightarrow v}/{A_u}$, i.e., the number of times $v$ has copied (e.g., retweeted) $u$, relative to the total activity (e.g., number of posts) of $u$ \cite{goyal2010learning}. Any edge with zero probability is removed, and the edges are normalized such that the sum of weighted out-degree for each node is 1.
        \item \textsc{Inf2vec}: 
         A shallow neural network performing IL based on the co-occurence of nodes in diffusion cascades and the underlying network \cite{feng2018inf2vec}. It was proven more effective than similar IL methods \cite{bourigault2016representation,saito2009learning} in the tasks of next node prediction and diffusion simulation.%
\end{itemize}

\subsection{Evaluation Methodology}
We split each dataset into train and test cascades based on their time of occurrence. The methods utilize the train cascades and/or the underlying network to define a seed set. The train cascades amount for the first 80\% of the whole set and the rest is left for testing.
The evaluation is twofold: computational time and seed set quality, similar to previous literature in influence maximization.
We evaluate the quality of the predicted seed set using  the number of \textit{Distinct Nodes Influenced} (DNI)\cite{complex2018,continest,panagopoulos2018diffugreedy}, which is the combined set of nodes that appear in the test cascades that are initiated from each one of the chosen seeds. To be specific, each predicted seed in the seed set has started some cascades in the test set. We compute the set of all infected nodes (DNI) by simply adding every node appearing in these test cascades, in a unified set. The size of this set indicates a measure of how successful was that seed set on the test set.
To understand the choice of DNI we need to take into consideration that each seed can create several test cascades. For example, another metric we could use is the sum of the average cascade length of each seed’s test cascades \cite{dynadiffuse}. With this approach though, we fail to counter for the overlap between different seeds’ spreads, and hence we can end up with influential seeds whose influence spreads overlaps a lot, resulting in an eventually smaller number of infected nodes. As mentioned above, DNI maintains a set of all distinct nodes participating in cascades started by each of the seeds, hence tackling the potential overlap issue. 
Overall, the most significant advantage over other standard IM evaluations is that it relies on actual spreading data instead of simulations.
Although not devoid of assumptions, it is the most objective measure of a seed set's quality, given the existence of empirical evidence. 

\begin{figure*}[t!]
\centering
\includegraphics[width=\textwidth]{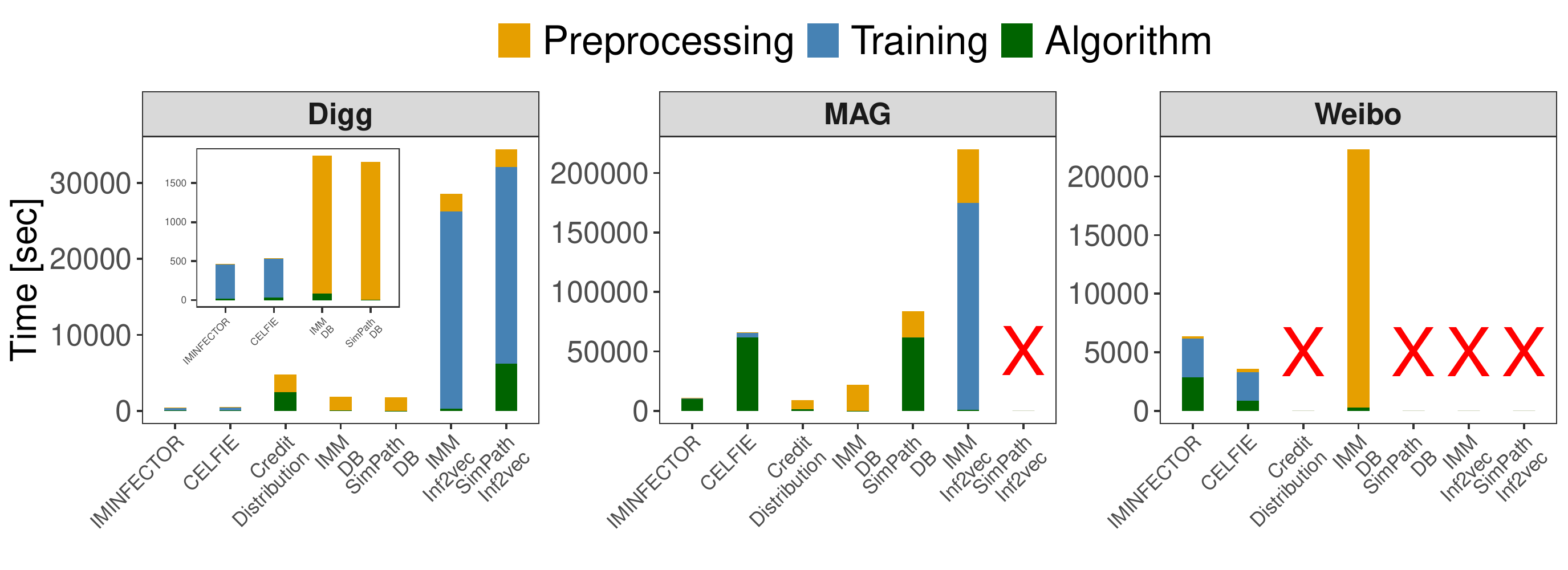}
\caption{Time comparison between the methods. Methods that could not scale are marked with \textcolor{red}{\textsf{X}}.}%
\label{fig:times}
\end{figure*}

\begin{figure*}[t!]
\centering
\includegraphics[width=1\textwidth]{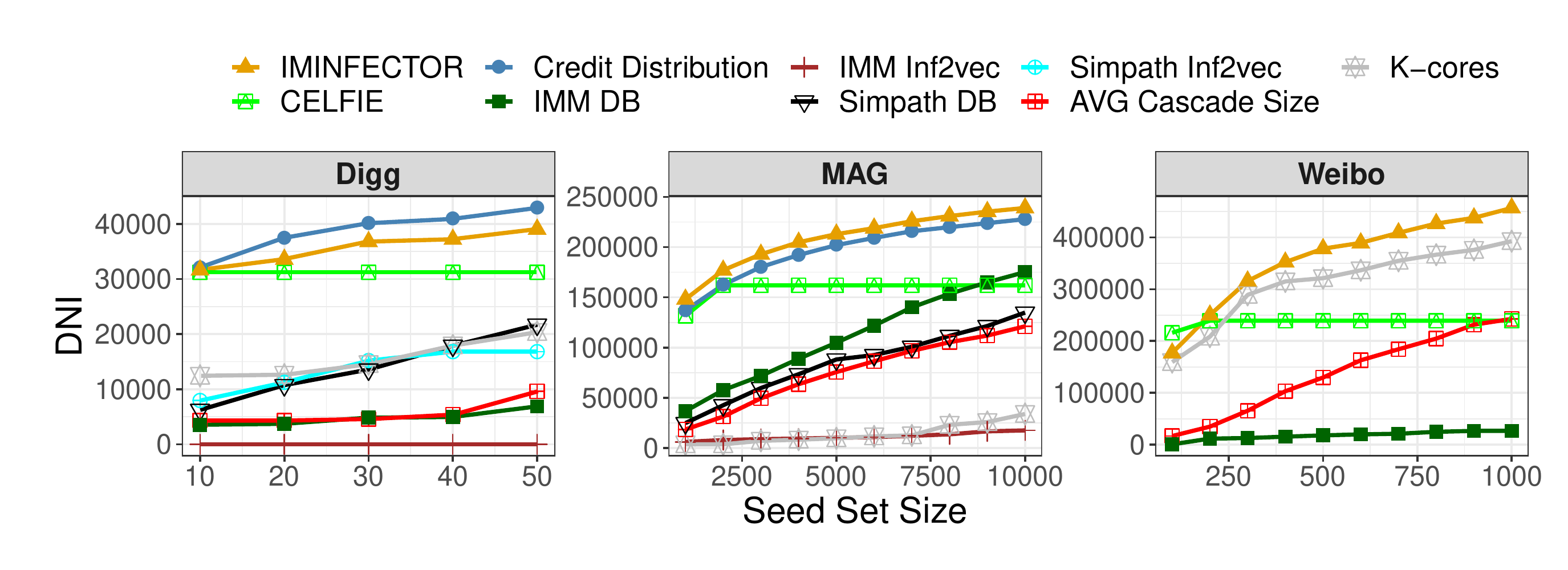}
\caption{The quality of the seed set derived by each method in different sizes, measured by the seed set's number of Distinct Nodes Influenced (DNI) in the test cascades. Algorithms that failed to scale for the minimum seed set size, are not included.}%
\label{fig:spreading}
\end{figure*}
Since our datasets differ significantly in terms of size, we have to use different seed set size for each one. For \textit{MAG} which has 205,839 initiators in the train set, we test it on 10,000, \textit{Weibo} with 26,158 is tested on 1,000, and \textit{Digg} with 537 has a seed set size of 50.
This modification is crucial to the objective evaluation of the methods because small seed sets favor simpler methods.
For example, the top 100 authors based on \textsc{K-cores} in \textit{MAG} would unavoidably work well because they are immensely successful. However, increasing the seed set size allows the effect of influence overlap to take place, and eventually, simplistic methods fall short. The experiments were run on a machine with Intel(R) Xeon(R) W-2145 CPU @ 3.70GHz, 252GB ram, and an Nvidia Titan V. Any method that required more memory or more than seven days to run, it is deemed unable to scale.

\subsection{Results}
Figure \ref{fig:times} shows the computational time of the examined methods, separated based on different parts of the model, and Figure \ref{fig:spreading} shows the estimated quality of each seed set. 
In terms of quality, we can see that \textsc{IMINFECTOR} surpasses the benchmarks in two of the three datasets. In \textit{Digg}, \textsc{Credit Distribution} performs better but is almost 10 times slower, because of the average cascade size in the dataset. 
All methods could scale in \textit{MAG} except for \textsc{SimPath} with \textsc{Inf2vec} weights, because in contrast to DB, \textsc{Inf2vec} retains all the edges of the original \textit{MAG} network, for which \textsc{SimPath} takes more than one week to run.  
Moreover, \textsc{SimPath} with both types of weights is not able to scale in the \textit{Weibo} dataset, due to the number of edges.
\textsc{Credit Distribution} also fails to scale in \textit{Weibo} due to the network's and cascade size. \textsc{IMM} with \textsc{Inf2vec} can not scale due to the high demand in memory. It can still scale with DB weights (close to 1M nodes), but it performs purely. 
The only method that scales successfully in Weibo was ranking by the $k$-core decomposition, which took about 650 seconds.
\textsc{IMM}'s performance is lower than expected, highlighting the difference between data-driven means of evaluation and the traditional simulations of diffusion models, which have been used in the literature due to the absence of empirical data. \textsc{IMM} optimizes diffusion simulations as part of its solution, so it is reasonable to outperform other methods in this type of evaluation. 
However, it performance is not equal in a metric based on real influence traces.
This can be attributed largely to the aforementioned problems of diffusion models and their miscalculation of the influence spread. 

Compared to CELFIE, the previous version of our algorithm, IMINFECTOR has superior performance as the seed set size increases. One of the CELFIE's core disadvantages is that the marginal influence spread, which is computed by sampling repetitively from a node's neighbors and summing the edge's weights, decreases very fast during the seeding rounds. This has as a result the whole network being covered with few rounds, and hence the retrieved seed set size to always be bounded.
In IMINFECTOR, each node has an expected influence spread as shown in Eq. (\ref{eq:exp_spread}),  which we have defined based on a node's influencer vector, compared to the rest of the initiator's representations. In other words, this formulation imposes the constraint that the network is shared among all candidate seeds, depending on their influencer vector, rendering infeasible to cover the whole graph without putting all initiators in the seed set.
Moreover, it allows for a fixed computation of a node's influence spread (e.g., without resampling), which renders it faster the CELFIE which retrieves only 127 seeds for Weibo, 2059 for MAG and 4 for Digg.

In terms of IL methods, DB outperformed \textsc{Inf2vec} because it removes edges with no presence in the cascades, diminishing significantly the size of the networks, some even close to 90\%.
However, it also had a severe shortcoming most notable in the social networks. 
Due to the general lack of consistent reposters and the huge amount of posts, the total activity of a node (the denominator in edge weight) was much larger than the number of reposts by a follower (numerator). Hence, the edge weights were too small with insignificant differences,  alleviating the computation of the simulated influence spread. 
This effect is not so strong in the bibliographical network, because authors tend to cite the same popular authors more consistently.
A side observation is the heavy computational burden of \textsc{Inf2vec}. It accounts for most of the computational time (blue color), which justifies the context creation mechanism of INFECTOR, while the accuracy validates our hypothesis that influencers' success lies more on the cascades they initiate rather their reposts.

In general, we see that \textsc{IMINFECTOR} provides a fair balance between computational efficiency and accuracy.
Most importantly, it exhibits such performance using \textit{only the cascades}, while the rest baselines use both, the network and cascades.
Being unaffected by the network size, \textsc{IMINFECTOR} scales with the average cascade size and the number of cascades,  which makes it suitable for real world applications where the networks are too big and scalability is of utmost importance.

\vspace{1cm}
\IEEEraisesectionheading{
\section{Conclusion} 
\label{sec:conclusion}}
In this paper, we have proposed \textsc{IMINFECTOR}, a model-independent method to perform influence maximization using representations learned from diffusion cascades. The machine learning part learns pairs of representations from diffusion cascades, while the algorithmic part uses the representations' norms and combinations to extract a seed set. The algorithm outperformed several methods based on a data-driven evaluation in three large scale datasets. 

It should be noted that our method is not devoid of assumptions. Firstly, we assume that the optimum seed set is comprised solely of nodes that initiate cascades. Given the task of influence maximization, this seems a valid assumption, with the exception of cases where very popular people in the real world enter the social network and get hyperconneccted without sharing content. Apart from this assumption, our method is highly scalable; a system though with too many cascade initiators (close to the number of users), will increase the overall complexity. A simple solution to this is to remove initiators with minuscule activity, which, given the exponential distribution characterizing the users' activity, will diminish severely the number. Finally, we tested our method in three datasets of varying sizes and characteristics, but there are further types of social networks that we need to experiment with to certify our method's effectiveness to the fullest.

\par A potential future direction is to examine deeper and more complex architectures for INFECTOR to capture more intricate relationships. 
From the algorithmic part, we can experiment with bipartite matching algorithms, given sufficient resources.
Overall, the main purpose of this work is primarily to examine the application of representation learning in the problem of influence maximization and secondarily to highlight the importance of data-driven evaluation (e.g., DNI). 
In addition, we want to underline the strength of model-independent methods and evaluations, given the recent findings on the severe shortcomings of diffusion models. 
We hope this will pave the way for more studies to tackle influence maximization with machine learning means.

 \bibliographystyle{IEEEtran}
 \bibliography{bib}
 

%

\appendices
\section{}
\subsection{Non-convex Influence Spread}
Given the diffusion matrix $\mathbf{D}$, the probability of a node $v$ getting influenced by a seed set $\mathcal{U}$ is the complementary probability of not getting influenced by each node $u \in \mathcal{U}$. Summing this over all non-seed nodes gives the number of nodes the $\mathcal{U}$ can influence:
\begin{equation}
    \sigma( \mathcal{U}) = \sum_{v \in G/\mathcal{U}}{\Big(1-\prod_{u \in \mathcal{U}}(1-p_{u,v})\Big)},
    \label{eq:new_spread}
\end{equation}
which is non-convex.
To transform this into a set function that computes the infected nodes, we can use a threshold, meaning that a node will get infected if its probability of getting influenced by the seed set at this step is equal to or more than 0.5, which is the value used in classifiers with softmax output.
Unfortunately, it is easy to see that this function is not submodular. A toy example with two source nodes $a, b$ that can influence three other nodes with probabilities $[0.6, 0.3, 0.5]$ and $[0.3,0.4,0.4]$, respectively. In the first step, the algorithm will choose $a$ as it gives 2 infected nodes opposing to $b$ that gives 0. In the second step, following our definition, the addition of $b$ will infect the second and final node, thus the property of diminishing returns does not stand for this influence spread and we can not utilize the greedy algorithm. Moreover the optimization of this non-convex function is non-trivial.

\subsection{Influence Spread by Bipartite Matching} \label{sec:appendix}
\begin{figure}[t]
\centering
\includegraphics[scale=.5]{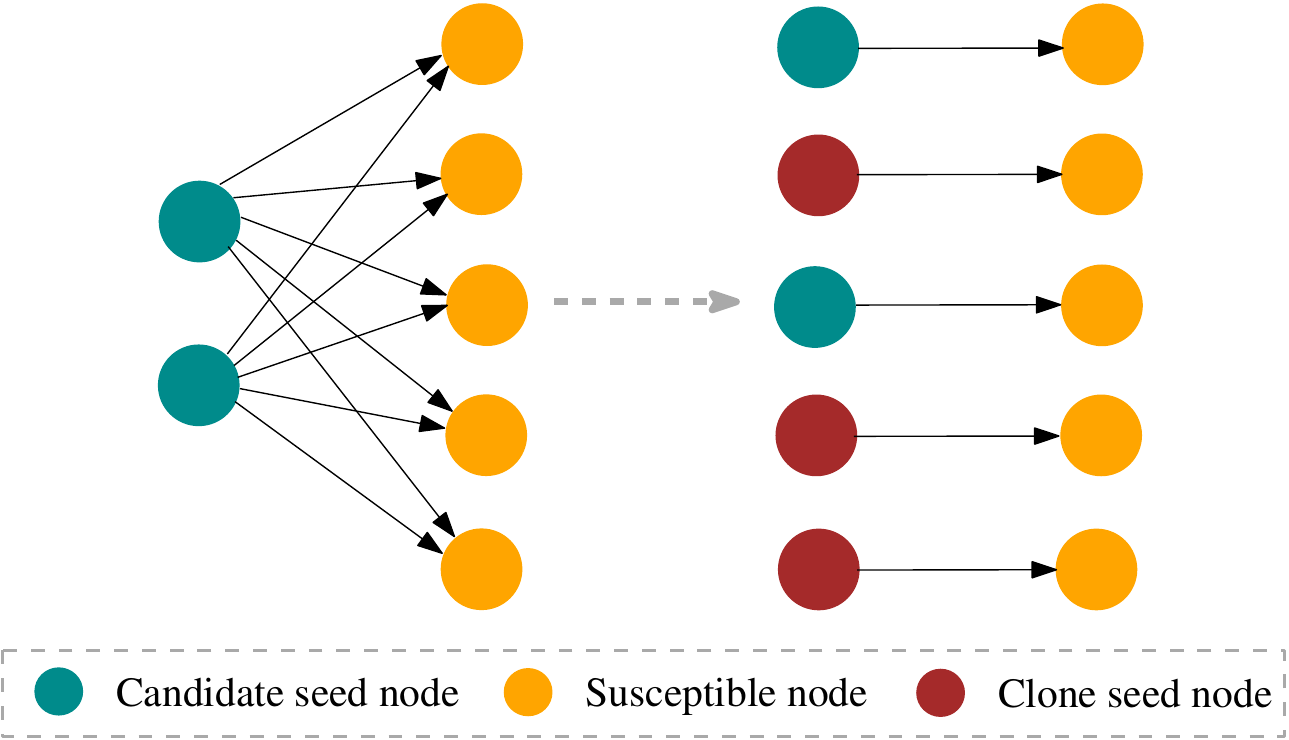}
\caption{Example of cloning candidate seeds for perfect bipartite matching.}
\label{fig:bipartite}
\end{figure}

The diffusion matrix $\mathbf{D}$ can be interpreted as a bipartite network with influencers (cascade initiators) on one side and susceptible nodes on the left. 
If we view the diffusion probabilities as edge weights, a simplification used extensively in similar context \cite{simpath,chen2010scalable}, we can transform the problem into weighted bipartite matching.
However, in this case, each candidate seed (left side) can be assigned to more than one nodes (right side). The matching can become perfect by creating a number of clone seeds for each candidate seed, and assigning them to the nodes the initial seed would be assigned to, in a one-to-one fashion, as can be seen in Figure \ref{fig:bipartite}.

The number of clones is basically equal to the expectation of a candidate seed $u$'s influence spread $\lambda_u$. Cloning each node $u$ for $\lambda_u$ times transforms $\mathbf{D}$ into a balanced, weighted bipartite graph, for which we could use algorithms such as the Munkres assignment \cite{munkres1957algorithms}. Unfortunately, it has a complexity of $\mathcal{O}(n^4)$ and requires constructing a non-sparse $N \times N$ matrix, which renders it prohibitive for real-world datasets. Though we can not use it directly, this interpretation remains useful for future directions.
\subsection{The Microsoft Academic Graph Dataset}
To construct the network and diffusion cascades from MAG, we utilized the tables PaperReferences, Author, PaperAuthorAffiliation,PaperFields and FieldsofStudy from the official from the official site\footnote{https://www.microsoft.com/en-us/research/project/microsoft-academic-graph/}.
Initially, we removed some one-paper authors based on a pattern we have identified in the past \cite{panagopoulos2019scientometrics}. Subsequently, we kept only papers with fields that are included in the "Computer Science" category. To create the coauthorship network, we formed a graph based on the authors' coappearence in these papers using the "PaperAuthorAffiliation" table. To form the diffusion cascades between authors, we had to derive a sequence of papers (citing other papers) using "Paper References". Subsequently, we transform it into a cascade over our network by substituting the papers with their authors, and forming one separate cascade for each author of the initial paper (as all of them can be considered initiators). The authors of a certain paper that is included in the cascade appear in a random sequence accompanied with the same timestamp, which corresponds to the time between the initial paper's publication and the time this paper was published.

\ifCLASSOPTIONcompsoc
\else
  \section*{Acknowledgment}
\fi

\ifCLASSOPTIONcaptionsoff
  \newpage
\fi



%

%

\begin{IEEEbiography}[{\includegraphics[width=1in,height=1.25in,clip,keepaspectratio]{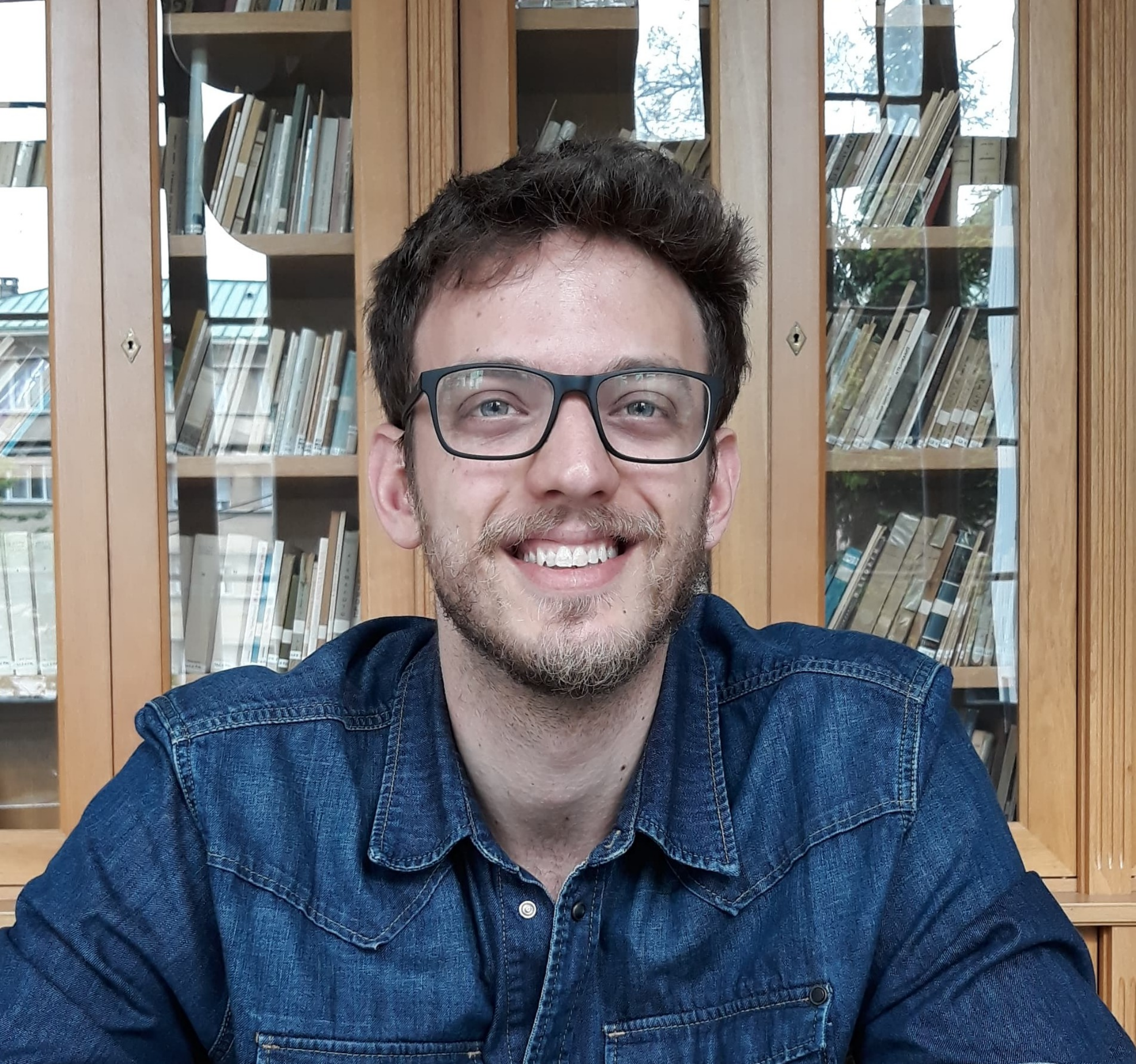}}]{George panagopoulos} is a PhD candidate at LIX Ecole Polytechnique, France.
He received his M.S. in Computer Science from University of Houston, US (2016), for which he received the best M.S. thesis award from the respective department.
At that time, he was a research assistant at the Computational
Physiology Lab and prior to that, at the Software Knowledge and Engineering Lab, NCSR Demokritos in Athens, Greece. 
His research interests lie in machine learning and its application to network science. 
\end{IEEEbiography}

\begin{IEEEbiography}[{\includegraphics[width=1in,height=1.25in,clip,keepaspectratio]{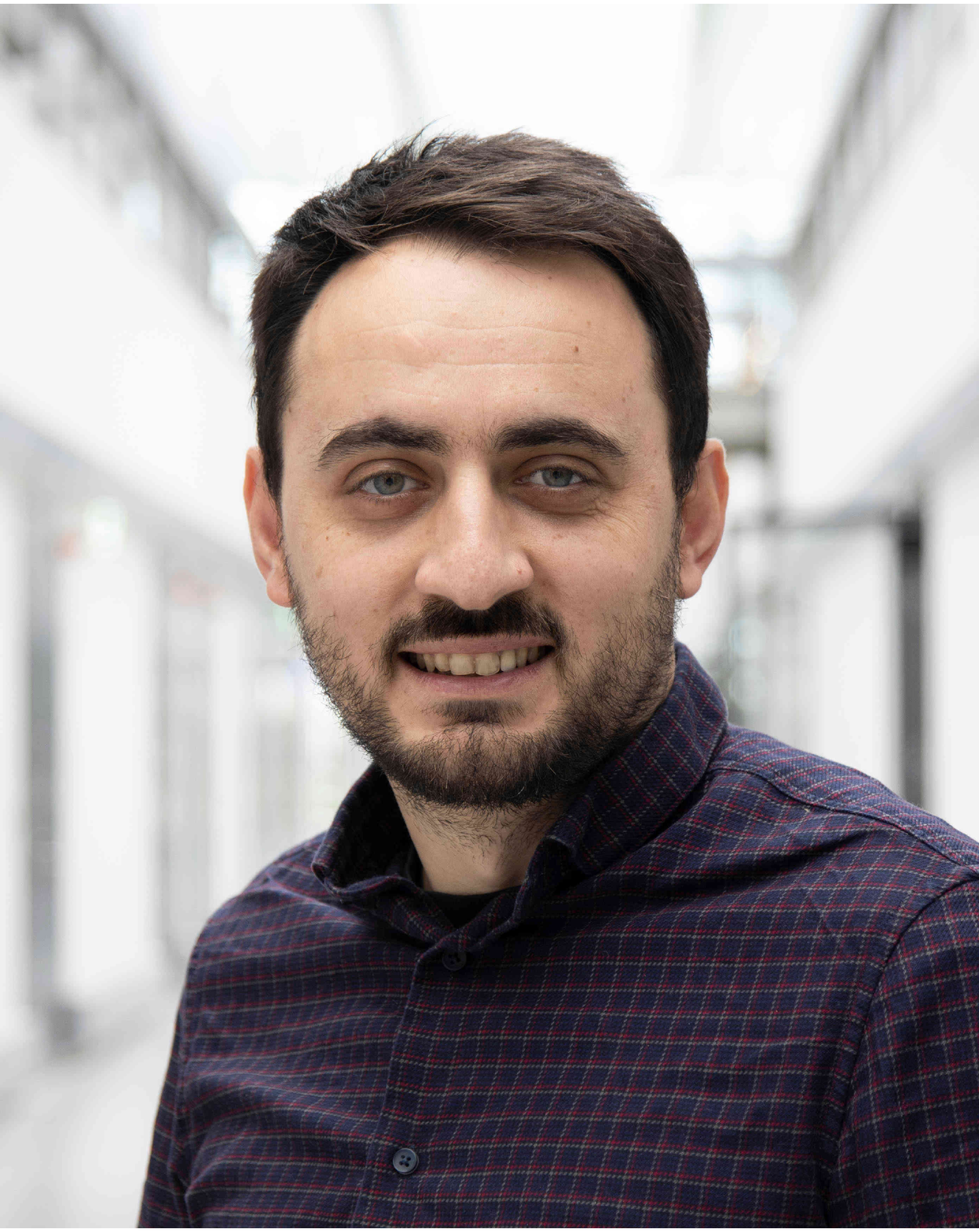}}]{Fragkiskos D. Malliaros} is an Assistant Professor at Paris-Saclay University, CentraleSupélec and associate researcher at
Inria Saclay. He also co-directs the M.Sc. Program in Data Sciences and Business Analytics (CentraleSupélec
and ESSEC Business School). Right before that, he was a postdoctoral researcher at UC San Diego (2016-17) and at
École Polytechnique (2015-16). He received his Ph.D. in Computer Science from École Polytechnique (2015) and his
M.Sc. degree from the University of Patras, Greece (2012). He is the recipient of the 2012 Google European Doctoral
Fellowship in Graph Mining and the 2015 Thesis Prize by École Polytechnique. In the past, he has been the co-chair of various data science-related workshops, and has also presented ten invited tutorials at international conferences in the area of graph mining and data science. His research interests span the broad area of data science, with focus on graph mining, machine learning and network analysis.
\end{IEEEbiography}

\begin{IEEEbiography}[{\includegraphics[width=1in,height=1.25in,clip,keepaspectratio]{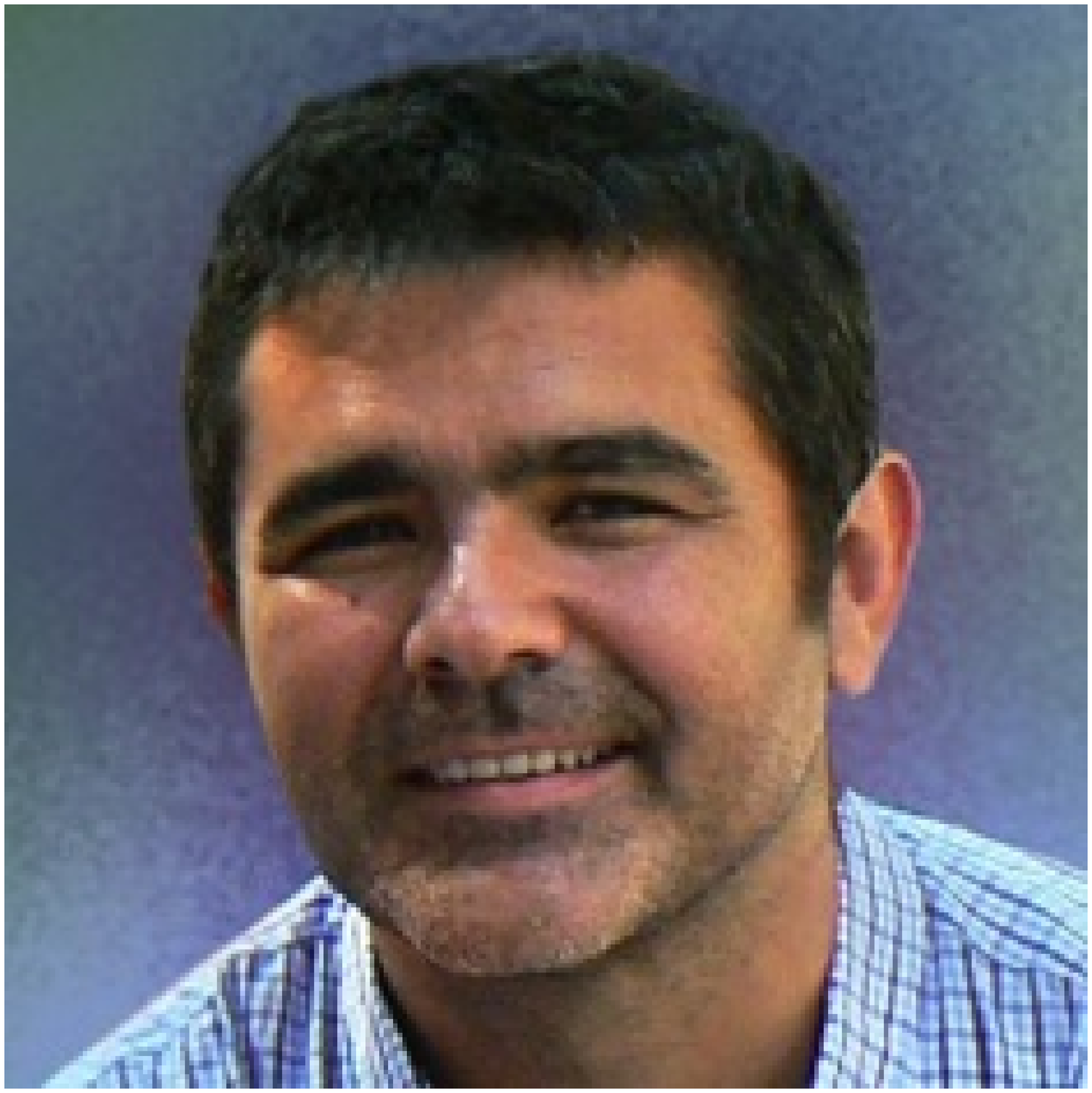}}]{Michalis Vazirgiannis} is a Professor at LIX, Ecole Polytechnique in France. He has conducted research in GMD-IPSI, Max Planck MPI (Germany), in INRIA/FUTURS (Paris). He has been teaching in AUEB (Greece), Ecole Polytechnique, Telecom-Paristech, ENS (France), Tsinghua (China) and in Deusto University (Spain). His current research interests are on machine learning and combinatorial methods for Graph analysis and Text mining. He has active cooperation with industrial partners in the area of data analytics and machine learning for large scale data repositories in different application domains such as Web advertising and recommendations, social networks, medical data, aircraft logs, insurance data. He has supervised fourteen completed PhD theses - several of the supported by industrial funding including Google and Airbus. On the publication frontier, he has contributed chapters in books and encyclopedias, published three books and more than a hundred forty papers in international refereed journals and conferences. He is also co-author of three patents. He has received the ERCIM and the Marie Curie EU fellowships and lead a DIGITEO chair grant. Since 2015 M. Vazirgiannis leads the AXA Data Science chair.
\end{IEEEbiography}





\end{document}